\documentclass[10pt,letterpaper]{article}
\usepackage[top=0.85in,left=1.0in,footskip=0.75in]{geometry}

\usepackage{amsmath,amssymb}

\usepackage{changepage}

\usepackage[utf8x]{inputenc}

\usepackage{textcomp,marvosym}

\usepackage{cite}

\usepackage{nameref,hyperref}

\usepackage[right]{lineno}

\usepackage{microtype}
\DisableLigatures[f]{encoding = *, family = * }

\usepackage[table]{xcolor}

\usepackage{array}

\newcolumntype{+}{!{\vrule width 2pt}}

\newlength\savedwidth

\raggedright
\usepackage[aboveskip=1pt,labelfont=bf,labelsep=period,justification=raggedright,singlelinecheck=off]{caption}

\makeatletter
\renewcommand{\@biblabel}[1]{\quad#1.}
\makeatother

\date{}

\usepackage{lastpage,fancyhdr,graphicx}

\newcommand{\sentzero}{$\mathit{Sent}(0)$}
\newcommand{\sentone}{$\mathit{Sent}(-1)$}

\usepackage{color}

\begin{document}
\vspace*{0.2in}

\begin{flushleft}
{\Large
\textbf\newline{Twitter Sentiment around the Earnings Announcement Events} 
}
\newline
\\
Peter Gabrov\v{s}ek\textsuperscript{\Yinyang},
Darko Aleksovski\textsuperscript{\Yinyang}*,
Igor Mozeti\v{c},
Miha Gr\v{c}ar
\\
\bigskip
Department of Knowledge Technologies, Jo\v{z}ef Stefan Institute, Ljubljana, Slovenia
\bigskip

\Yinyang These authors contributed equally to this work.
\bigskip

* darko.aleksovski@ijs.si (DA)

\end{flushleft}


\section*{Abstract}
We investigate the relationship between social media, Twitter in particular,
and stock market. We provide an in-depth analysis of the Twitter volume and
sentiment about the 30 companies in the Dow Jones Industrial Average
index, over a period of three years. We focus on Earnings Announcements
and show that there is a considerable difference with respect to when the
announcements are made: before the market opens or after the market closes.
The two different timings of the Earnings Announcements were already investigated 
in the financial literature, but not yet in the social media.
We analyze the differences in terms of the Twitter volumes, cumulative
abnormal returns, trade returns, and earnings surprises.

We report mixed results.
On the one hand, we show that the Twitter sentiment (the collective opinion 
of the users) on the day of the announcement very well reflects the stock moves 
on the same day.
We demonstrate this by applying the event study methodology, where
the polarity of the Earnings Announcements is computed from the Twitter sentiment.
Cumulative abnormal returns are high (2--4\%) and statistically significant.
On the other hand, we find only weak predictive power of the Twitter sentiment
one day in advance.
It turns out that it is important how to account for the announcements made 
after the market closes. 
These after-hours announcements draw high Twitter activity immediately,
but volume and price changes in trading are observed only on the next day.
On the day before the announcements, the Twitter volume is low, and the sentiment 
has very weak predictive power.
A useful lesson learned is the importance of the proper alignment between
the announcements, trading and Twitter data.


\section{Introduction}

It is now accepted that financial markets are not governed solely by
rational behavior of investors, as captured by the efficient market hypothesis.
Their decisions are also influenced by their subjective beliefs and expectations,
and by the information from the Internet. Online news and social media
provide large amounts of data, from which potentially useful information
can be extracted. We are interested in collective opinion and expectations
of investors in relation to financial markets. We analyze
social media data from the Twitter micro-blogging platform in terms of
the attention to the most important events, and the collective expectations
about the market moves.

Twitter is becoming an increasingly popular platform used
to monitor and forecast financial markets.
The first to show a clear relation between the Twitter
mood indicators and Dow Jones Industrial Average (DJIA) were 
Bollen et al. \cite{bollen2011twitter, bollen2011modeling, mao2011predicting}.
In general, related work provides mixed conclusions about the 
relation between Twitter and stock markets. The results depend on the
type of analyses performed and whether the individual stocks
or aggregate indices are considered.
The work of Preis et al. \cite{preis2010complex} studies the relation over 
time between the daily number of search queries for a particular stock and 
the volume of daily trades with the same stock. 
The number of search queries is also analyzed as a proxy for the popularity 
of stocks and stock riskiness \cite{kristoufek2013can}.

The range of the methodologies for analyzing the relations consists of
the Granger causality \cite{mao2011predicting, bollen2011twitter, Souza2015}, 
one-step ahead forecasting analyses 
\cite{bollen2011twitter, mao2011predicting, wang2014crowds, Souza2015},
information theoretic approaches \cite{zheludev2014can, Souza2016},
and event studies \cite{Sprenger2014jbfa, Sprenger2014efm, Ranco2015eventstudy}.
On the one hand, Granger causality and the information theoretic 
approaches analyze time series over a longer time period. They 
provide results about the existence of a quasi-causal relation between the 
social media and stock market. They do not identify precise time periods
when this relation is stronger, weaker, or non-existent. 
Event studies, on the other hand, focus on the relations in specific 
time intervals, thus providing potentially more actionable evidence for 
trading purposes.

Several papers analyze the relation between Twitter and stock market
for the aggregated indices only, e.g., 
DJIA \cite{bollen2011twitter,mao2011predicting}
or S\&P 500 \cite{mao2012correlating, makrehchi2013stock, wang2014crowds}.
Only a few provide conclusions regarding the relation between the Twitter posts
and stocks of individual companies, e.g., 
\cite{zheludev2014can, Sprenger2014jbfa, Sprenger2014efm, Ranco2015eventstudy}.
The main reason is the typically insufficient number of Twitter posts
about individual companies to draw statistically significant results.

We overcome this limitation by focusing on the most interesting time periods
of an individual company, the quarterly Earnings Announcements (EA).
It turns out that the volume of Twitter posts around most of
the EAs is substantially higher and allows to draw
statistically significant conclusions.
This type of analysis is enabled by the ``event study'' methodology
\cite{mackinlay1997event, campbell1997econometrics}, used in economics and finance.
The event study has been often used to verify if the content of EAs
conveys useful information for the valuation of companies.
It allows to draw conclusions about the price movement of a stock
on average, over several different events of the same type. In related
work in economics, the event study typically relies on the earnings surprise, i.e., 
the difference between the expectations of financial analysts and the reported
valuation of a company in its earnings report. In our work, however, we test 
if the aggregate sentiment expressed in financial tweets
around the EAs indicates the direction of the stock price movement.

There have been applications of the event study methodology to
Twitter data already. We are aware of three recent works which analyze
the Twitter sentiment data in relation to the stock price movement.
Sprenger et al. \cite{Sprenger2014jbfa, Sprenger2014efm} analyze
known EA events as well as other, unexpected events. They conclude that 
both the sentiment and the type of the news can explain the market 
reaction (movement of the stock price). 
Our previous work by Ranco et al. \cite{Ranco2015eventstudy}
presents evidence of significant dependence between stock price returns
and Twitter sentiment in tweets about the companies.
Ranco et al. apply an event detection procedure to detect events from the Twitter data.
They report that most of the EA events have a corresponding peak in the Twitter volume.
This fact is the main motivation for the current work, where we focus on and
perform an in-depth analysis of the EA events.

There are several improvements over our previous work \cite{Ranco2015eventstudy}.
In this study we analyze data over a longer period of three years.
There are over 4.5 million tweets labeled with sentiment, one of the
largest datasets available.
We provide and analyze Twitter data at hourly resolution, which enables fine-tuned 
aggregation of tweets at the daily resolution.
Instead of analyzing events detected from the Twitter stream, we focus only on 
the EA events, which are known in advance.
We observe important differences between the timings of the EA events,
either before the market opens, or after the market closes.
It turns out that it is crucial to appropriately aggregate the tweets at
daily resolution and align them with the market activities.
Finally, we use a different, formally sound, sentiment measure than the one used 
by Ranco et al. \cite{Ranco2015eventstudy}.

There are several works in the financial literature, dealing with EAs and their
timings, related to our work. Berkman et al. \cite{berkman2009event} 
observe that the proportion of EAs reported after the close of the market has increased
in recent years. Their main conclusion is that it is important to account for the 
after-hours announcements when performing event studies. Specifically, in the case of 
an after-hours announcement, the day 0 prices should be shifted to the next market day.
In case that no shifting is performed, abnormal returns could be biased.

Doyle et al. \cite{doyle2009timing} find that typical companies consistently 
report either after the close of the market, or before the market opens. 
They conclude that more complex companies tend to announce after the market closes. 
They also find higher trading volume around the after close announcements in
comparison to the before open announcements --- our study confirms the same observation.
These two facts provide evidence for their hypothesis that reporting after the market closes
allows for broader dissemination of the information contained in the announcements.

Schroff et al. \cite{schroff2013individual} analyze the collective actions of
individual investors around EAs. They find that individual investors
take more risk and tend to overestimate the likelihood of positive events while 
underestimating the likelihood of negative ones.
This intense buying (selling) actions of individual investors prior to
EAs are followed by significant positive (negative) abnormal
returns after the events \cite{kaniel2007investor}.

A recent work by Alostad et al. \cite{alostad2015} is closely related to the event study 
applied here. They combine two types of complementary data: volume from Twitter and 
sentiment from financial news. They conclude that it is useful to predict the 
direction of a stock price move only when there is an abnormally high Twitter volume.
However, in contrast to the other event studies \cite{Ranco2015eventstudy, Sprenger2014jbfa},
they assign polarity of the events from the financial news instead of the tweets.
The work of Tafti et al. \cite{tafti2016realtime} relates the peaks of Twitter volume 
(events detectable from the Twitter time series) to the subsequent increase 
of trading volume. The authors conclude that it is difficult to make use of the 
information from Twitter for forecasting purposes.

This work makes several contributions to the analysis of
relations between social media and stock market.
First, we find significant relation between the Twitter sentiment and EA returns
on the days of the announcements, with cumulative abnormal returns around 2--4\%.
Second, we observe important differences between different timings of the announcements.
The announcements before the market opens show lower cumulative abnormal returns
in comparison to the announcements after the market closes.
Third, based on these results, we test a simple trading strategy with buy/hold/sell at the
market close of the day prior to the EA. 
Fourth, we compare the Twitter sentiment to the earnings surprise, 
a measure frequently used in event studies.

An important conclusion from these results is the requirement
for a proper setup of event studies and other methods which investigate
relations between Twitter and stock market.
It is important to take into account the exact timing of the events
with respect to the market trading hours.
The daily aggregation of the Twitter data and its alignment 
with the market data have an impact on the perceived predictive power
of Twitter.
When analyzing the data at the daily resolution
one needs to attribute the Twitter posts created after the market closes 
to the next trading day.



\section{Data}

In this section we provide details about the data used in the study.
The data are about the 30 companies in the DJIA index in the period
of three years, from June 1, 2013, to June 3, 2016.
The data consist of the Earnings Announcement (EA) events,
the stock market data, and the Twitter data.
All the data were collected according to the Terms of Use and Service of 
the source websites, and are available at
\url{https://dx.doi.org/10.6084/m9.figshare.4036269}.
The terminology and abbreviations used throughout the paper are in
Table \ref{tab:terminology}.
 
\begin{table}[!h]
\centering
\caption{{\bf Terminology and abbreviations used.}}
\label{tab:terminology}
\begin{tabular}{cl}
\hline
DJIA & Dow Jones Industrial Average index \\
NYSE & New York Stock Exchange \\
EA & Earnings Announcement \\
Day 0 & the day of the EA \\
Day $-1$ & the day before the EA \\
CAR & Cumulative Abnormal Return \\
EPS & Earnings Per Share \\
ES  & Earnings Surprise \\
BeforeOpen & EA before the opening of NYSE (9:30 a.m. US/Eastern)\\
AfterClose & EA after the closing of NYSE (4:00 p.m. US/Eastern)\\
Twitter volume & number of tweets at a given resolution \\
Twitter sentiment & stance or leaning w.r.t. stock price move \\
\sentzero & Twitter sentiment on the day of the EA \\
\sentone & Twitter sentiment on the day before the EA \\
\hline
\end{tabular}
\end{table}

\subsection{Earnings announcement data}

The data regarding the EAs contains the exact timings 
of the announcements, as well as the reported and expected price of a share.
The earnings surprise is the difference between the reported and 
the expected earnings of a company. The earnings surprise \textit{ES} is defined as:
\begin{equation}
ES = \frac{p_{rep}-p_{est}}{p_{est}}
\label{eq:eps}
\end{equation}
where $p_{rep}$ is the reported price of a share in the EA report, 
while $p_{est}$ is the expected price, as estimated by financial analysts.
We collected the EA data from the \url{http://www.zacks.com} website. 
The missing values of the timings were filled from the information issued by the 
companies themselves.

\subsection{Market data}

The market data was collected from \url{http://www.google.com/finance}.
It consists of the daily trading volume and closing prices 
of the 30 DJIA companies and the DJIA index. 
From the data we calculate daily returns and longer term trading returns.
The daily return $R_d$, used in the calculation of Cumulative Abnormal
Returns (event study applications), is defined as:
\begin{equation}
R_d=\frac{p_{d}-p_{d-1}}{p_{d-1}}
\label{eq:trade_return}
\end{equation}
where $p_{d}$ denotes the closing price of the stock on day $d$.
Consistent with the original event study \cite{mackinlay1997event},
we operate with raw-returns, and not the more standard log-returns.

The longer term trading return, on the other hand, compares prices over several days, 
and is used in the evaluation of trading strategies. We take as a basis the closing
price of a stock on the day before the EA (day $-1$) and compare it
to the closing price after the EA. 
The trading return $RT_{d}$ is defined as:
\begin{equation}
RT_{d} = \frac{p_{d} - p_{-1}}{p_{-1}}
\label{eq:trade_return_trading}
\end{equation}
where $p_{-1}$ is the closing price on the day prior to the EA (day $-1$),
and $p_d$ is the closing price on the trading day $d$ after the EA. 
Note that the trading return is computed as the relative difference in
prices over $d+1$ days.

\subsection{Twitter data}

The Twitter data used in this study is summarized in Table \ref{tab:tweets}
and contains approximately 4.5 million tweets for 30 companies, 
during a period of three years.
The data were collected by the Twitter Search API, where a query is specified
by the stock cash-tag (e.g., ``\$MSFT'' for Microsoft).

\begin{table}[!h]
\centering
\caption{{\bf The data about the 30 DJIA companies.} 
The collected tweets and Earnings Announcements (EA) cover the period 
of three years, from June 1, 2013 to June 3, 2016.
Companies are ordered by the total number of tweets collected.
For each company, there is the sentiment distribution, market capitalization,
and the prevailing timing of EAs with respect to the NYSE trading hours.
Each company issues four EAs per year, therefore there is a total of 360 EAs
(30 companies, three years, four EAs per year)$^{1}$.
}
\label{tab:tweets}
\begin{tabular}{ll|rrrr|rl}
\hline
       &                                     &  \multicolumn{4}{c|}{Number of tweets} & Market cap & Earnings \\
Ticker &                             Company &  Negative &  Neutral &  Positive &  Total & [$10^9$ US\$] & Announcements \\
\hline
  MSFT &                      Microsoft Corp &     31,626 &   328,336 &     72,961 &  432,923 & 449.39 & AfterClose \\
   IBM &  			Intl. Business Machines  &     26,318 &   204,219 &     38,685 &  269,222 & 152.50 & AfterClose$^2$ \\
    GS &             Goldman Sachs Group Inc &     24,708 &   205,860 &     34,005 &  264,573 & 72.14 & BeforeOpen \\
   JPM &               JPMorgan Chase and Co &     35,263 &   183,407 &     32,395 &  251,065 & 243.74 & BeforeOpen$^{1,3}$ \\
   DIS &                      Walt Disney Co &     15,060 &   166,657 &     43,968 &  225,685 & 151.74 & AfterClose \\
  INTC &                          Intel Corp &     16,222 &   156,301 &     37,267 &  209,790 & 170.69 & AfterClose \\
     T &                           AT\&T Inc &     10,039 &   156,935 &     28,113 &  195,087 & 251.92 & AfterClose \\
    GE &                 General Electric Co &      9,285 &   157,059 &     27,477 &  193,821 & 280.40 & BeforeOpen \\
   WMT &                 Wal-Mart Stores Inc &     21,774 &   141,010 &     25,767 &  188,551 & 224.26 & BeforeOpen \\
   XOM &                    Exxon Mobil Corp &     17,864 &   140,413 &     24,406 &  182,683 & 362.50 & BeforeOpen \\
  CSCO &                   Cisco Systems Inc &     11,822 &   125,233 &     29,763 &  166,818 & 160.14 & AfterClose \\
   MCD &                     McDonald's Corp &     19,554 &   121,320 &     21,612 &  162,486 & 98.84 & BeforeOpen \\
   PFE &                          Pfizer Inc &      7,541 &   115,453 &     24,422 &  147,416 & 210.90 & BeforeOpen \\
   JNJ &                   Johnson \& Johnson &     10,700 &   108,927 &     20,888 &  140,515 & 326.44 & BeforeOpen \\
    KO &                        Coca-Cola Co &     10,156 &   105,703 &     21,851 &  137,710 & 188.44 & BeforeOpen \\
   MRK &                      Merck \& Co Inc &      5,826 &   101,404 &     18,375 &  125,605 & 174.15 & BeforeOpen \\
   CAT &                     Caterpillar Inc &     15,611 &    91,688 &     16,480 &  123,779 & 47.73 & BeforeOpen \\
     V &                            Visa Inc &      7,456 &    94,786 &     21,047 &  123,289 & 193.52 & AfterClose$^4$ \\
   NKE &                            Nike Inc &      8,431 &    83,267 &     31,308 &  123,006 & 97.34 & AfterClose \\
   CVX &                        Chevron Corp &     11,911 &    90,240 &     17,399 &  119,550 & 190.41 & BeforeOpen \\
    BA &                           Boeing Co &     11,090 &    82,097 &     24,410 &  117,597 & 81.82 & BeforeOpen \\
    VZ &              Verizon Communications &      7,200 &    86,632 &     20,917 &  114,749 & 215.55 & BeforeOpen \\
    HD &                      Home Depot Inc &      7,091 &    74,311 &     20,701 &  102,103 & 166.99 & BeforeOpen \\
   AXP &                 American Express Co &      7,378 &    64,912 &     11,665 &   83,955 & 60.22 & AfterClose \\
    PG &                 Procter \& Gamble Co &      6,393 &    63,960 &     12,241 &   82,594 & 235.38 & BeforeOpen \\
   UNH &              UnitedHealth Group Inc &      4,596 &    42,602 &      9,817 &   57,015 & 130.11 & BeforeOpen \\
    DD &                             DuPont  &      4,400 &    43,164 &      7,887 &   55,451 & 61.28 & BeforeOpen \\
   MMM &                               3M Co &      4,020 &    40,262 &      8,485 &   52,767 & 109.28 & BeforeOpen \\
   UTX &            United Technologies Corp &      3,652 &    31,293 &      7,725 &   42,670 & 89.49 & BeforeOpen \\
   TRV &            Travelers Companies Corp &      2,772 &    18,649 &      4,526 &   25,947 & 34.26 & BeforeOpen \\
\hline
 Total &                                     &    375,759 & 	3,426,100 &  	 716,563 & 4,518,422 & 5,231.57 & 359 \\
\hline
\end{tabular}
\begin{flushleft}
$^1$ There are no tweets on the day of one EA - as a consequence, we consider the total of 359 EAs instead of 360. \\
Exceptions to the prevailing timings of EAs: \\
$^2$ all EAs are AfterClose, except one is BeforeOpen, \\
$^3$ all EAs are BeforeOpen, except one is AfterClose, \\
$^4$ all EAs are AfterClose, except two are BeforeOpen.
\end{flushleft}
\end{table}

Each tweet is then labeled with a `sentiment' with three possible values: 
negative, neutral, or positive. The label denotes the future stock price move, as
anticipated by the author of the tweet: down (negative sentiment),
unchanged (neutral), or up (positive). 
The term `Twitter sentiment' used here is misleading and used for historical
reasons. What is actually meant is stance \cite{Mohammad2016} or leaning 
of a Twitter user w.r.t. the future stock price move. The sentiment vocabulary alone, 
positive or negative, used in a tweet does not necessarily reflect the
user expectations about the stock price move, therefore all the relevant vocabulary
is explored. The tweets were labeled for
sentiment automatically, by a supervised learning method, 
described in more detail in the ``\nameref{subsect:sent_class}'' subsection
in ``\nameref{sec:methods}''.

The tweets about each company are aggregated at hourly and daily resolution.
The close of the market is used to delimit the daily aggregation of tweets.
The sentiment of a set of tweets at day $d$ is defined by the sentiment score
$\mathit{Sent}_d$:
\begin{equation}
Sent_d = \frac{N_d(pos) - N_d(neg)}{N_d(pos) + N_d(neut) + N_d(neg) + 3}
\label{eq:sentiment_score}
\end{equation}
where $N_d(neg)$,  $N_d(neut)$ and $N_d(pos)$ denote the daily number of 
negative, neutral and positive tweets, respectively. The sentiment
score has the range $-1 < Sent < +1$.
Formally, the sentiment score is the mean of a discrete probability
distribution with values of $-1$, $0$ and $+1$ for negative, neutral
and positive sentiment, respectively \cite{Kralj2015emojis}. The probabilities of 
each label are estimated from their relative frequencies, but when dealing 
with small samples (e.g., only a few tweets about a stock per day) it is 
recommended to estimate probabilities with \textit{Laplace estimate}. 
This is the reason for the constant $3$, the number of discrete labels, 
in the denominator of Eq (\ref{eq:sentiment_score}).

Each company in Table \ref{tab:tweets} is identified by its ticker symbol 
according to the NYSE. The table also reports the number and 
sentiment distribution of tweets, the market capitalization, and the timing
used by the company to report the EAs. Most of the companies time their reports 
consistently, either always AfterClose or BeforeOpen. There are only a few 
exceptions which switch between the two options \cite{truong2010strategic}.
In summary, out of a total of 359 EAs during the three years there are 
253 BeforeOpen announcements, and 106 AfterClose announcements.

\subsection{Data alignment}

We focus on the relations between the stock market and Twitter posts around
the EAs. Of particular importance for our analyses is proper alignment of 
the Twitter and market data.

\begin{figure}[!h]
\centering
\includegraphics[width=\textwidth]{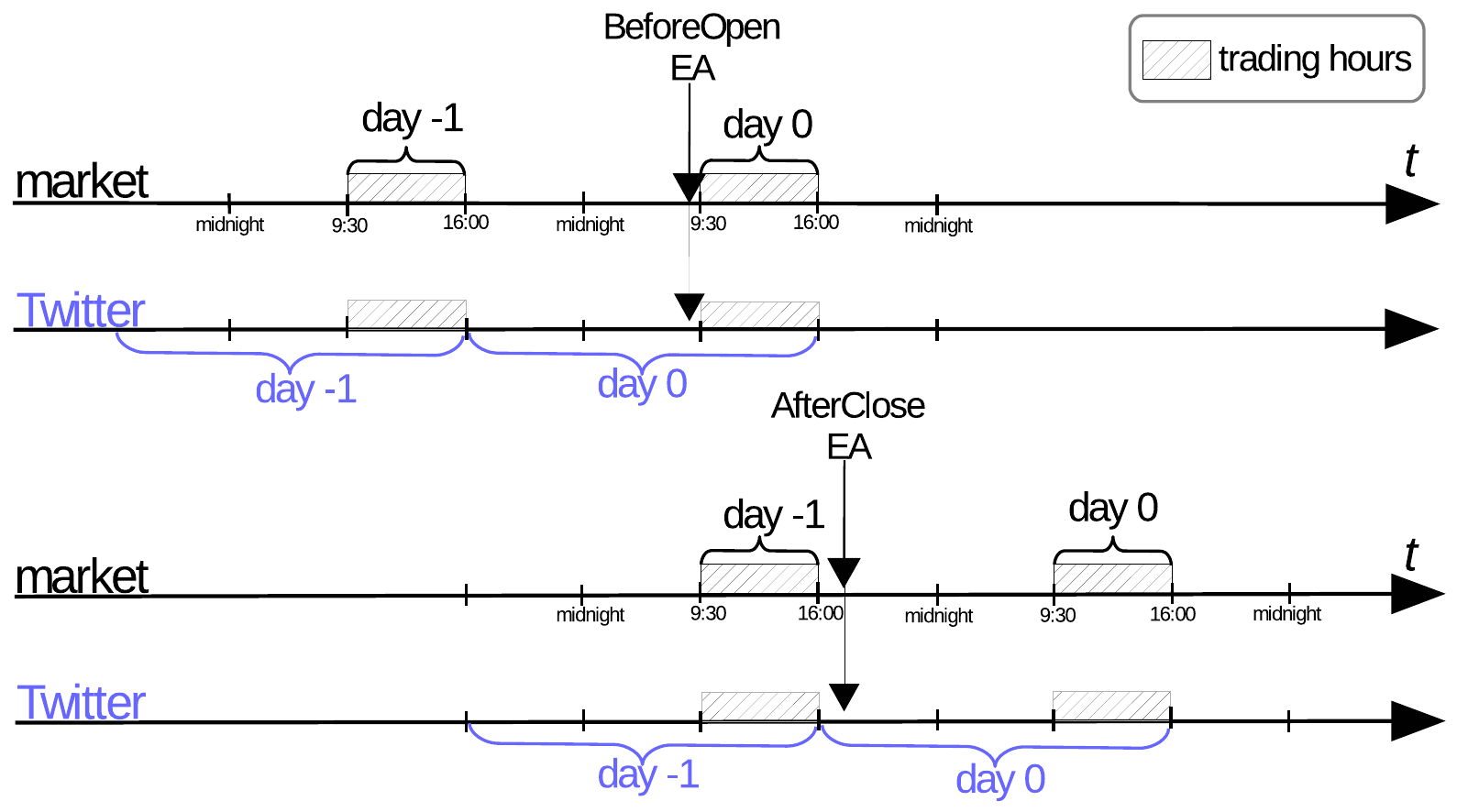}
\caption{{\bf Timings of the Earnings Announcements.}
Relation between the two different types of EAs, trading hours and 
daily aggregation of tweets.}
\label{fig:timeline}
\end{figure}

We focus on the days of the
EAs (denoted as day $0$), and the days immediately before the EAs (day $-1$).
There is an important distinction when exactly are the EAs made
with respect to the NYSE trading hours. Some announcements (denoted
BeforeOpen) are made before the market opens (9:30 a.m. US/Eastern),
and some (denoted AfterClose) are made after the market closes (4:00 p.m. US/Eastern).
Fig \ref{fig:timeline} depicts the relation between the trading hours and EAs. 
Note that the days are delimited by the market close hour,
and not by the midnight. Consequently, the day $0$ trading for the 
AfterClose announcements occurs on the next calendar day
(see the lower part of Fig \ref{fig:timeline}).
This is consistent with the treatment of the
EAs in the financial literature \cite{berkman2009event}.

\section{Results and Discussion}

This section presents analyses of interactions between social
media (Twitter) and financial market (stock returns).
First we compare the volume of tweets to the trading volume
and notice very similar patterns in both systems, namely,
highly elevated activity around the EA events.
We observe a considerably higher trading activity for the
AfterClose announcements, and find similar results for the
Twitter volume at hourly resolution.

Next, we adapt and apply the event study methodology to our data.
Event study, as defined in financial econometrics 
\cite{mackinlay1997event, campbell1997econometrics},
analyses abnormal stock returns during external events.
The external events are first identified and grouped into categories
whether the event should have positive, negative, or no effect on the stock returns.
The null hypothesis $H_0$ is that external events have no impact on the returns.
Under $H_0$, abnormal returns are normally distributed \cite{campbell1997econometrics},
therefore one can test whether abnormal returns during external events are
statistically significant.

In our study, the EAs are the only external events we consider.
We derive the categorization of the EAs (negative, neutral, or positive)
from the Twitter sentiment alone,
and not from the EA reports, as in the standard approaches.
We test the $H_0$ for the AfterClose and BeforeOpen announcements
separately, and find very different results.
We are not only interested in the significance of abnormal returns,
but also in their magnitude.
Further, we test if the Twitter sentiment has any predictive power,
i.e., if the EA reports are anticipated in the social media before the
actual announcements are made.

We compare Cumulative Abnormal Returns (from the event study) with
trading returns of the stocks and the DJIA index, and find very similar results.
Based on these, we propose and backtest a simple trading strategy
over the period of three and half years.

Finally, we compare the relation between the Twitter sentiment score to
the earnings surprise values. We find very weak relation between the two variables,
but we observe some difference between the AfterClose and BeforeOpen announcements.

\subsection{Twitter and trading volumes}

The goal of this subsection is to analyze the activities on social media
around the Earnings Announcements. If one observes an elevated Twitter activity
together with higher trading activity than this indicates that the EA events
are reflected in social media. This is a motivation to apply the event
study methodology, described in the next subsection, to analyze if there 
are also abnormal returns corresponding to the sentiment signal from Twitter. 

\begin{figure}[!h]
\centering
\includegraphics[width=\textwidth]{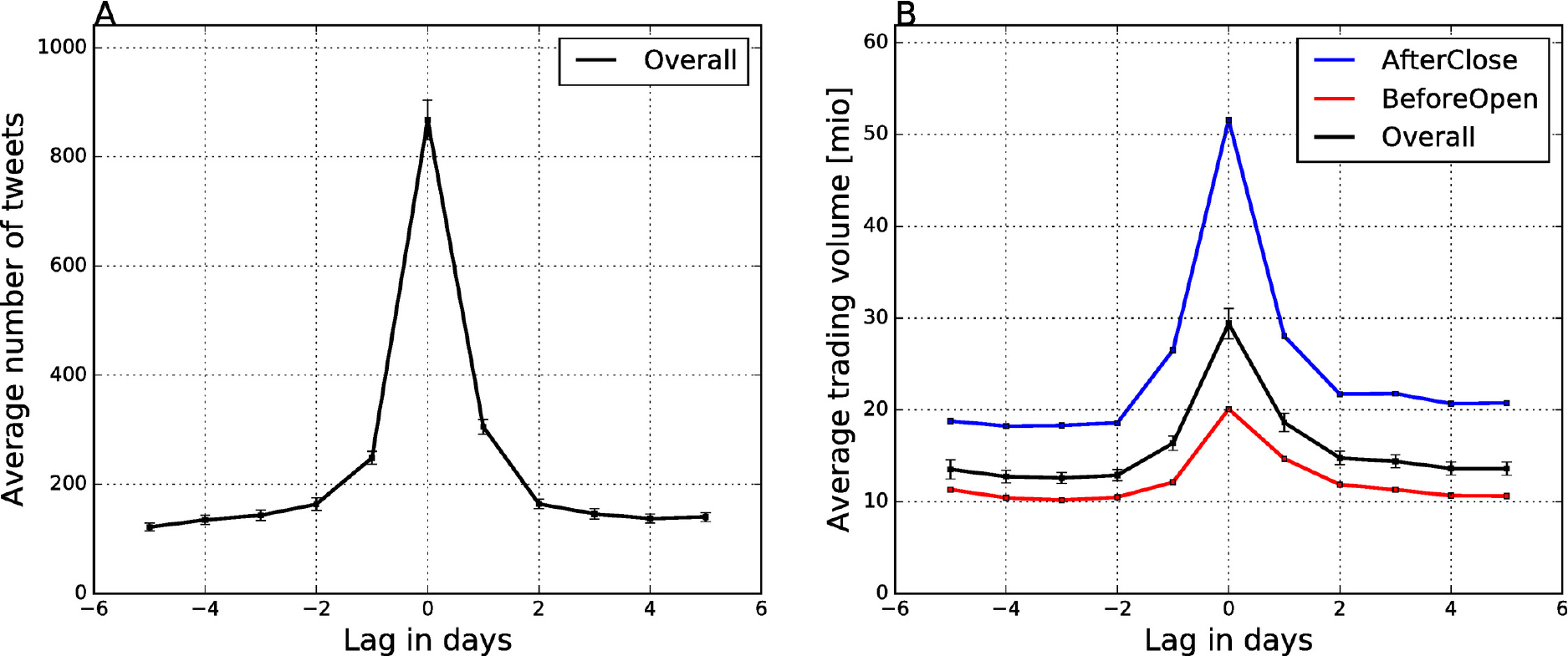}
\caption{{\bf Daily number of tweets (A) and trading volume (B) around 
the Earnings Announcements.}
The overall average number of tweets per trading day is 200.
The trading volume (B) shows the overall average across all EAs (black line),
the average trading for the AfterClose (blue line), and for the BeforeOpen 
(red line) announcements. Error bars around the black lines denote
one standard error.}
\label{fig:daily_volume}
\end{figure}

We first compare the activity of Twitter users and stock traders on the days 
around the EAs. The Twitter activity is estimated by the average number of tweets
per day, and the trading activity by the average daily trading volume.
The results are in Fig \ref{fig:daily_volume}.

We consider five days around the day of an EA (day $0$) and
observe a very similar pattern of elevated activity in both cases.
The average number of tweets over the three years is 200 tweets per trading day.
The above average activity is observed not only on day $0$, but also
on days $-1$ and $+1$. Cumulatively, the three days around the EAs
exhibit $2.4$ larger volume than on the average.
This indicates that Twitter users are active around the EAs days and post
their opinions about the companies and their finances.

The trading activity around the EAs is also higher, since the trading volume increases
in the same time period, as shown by the black line in Fig \ref{fig:daily_volume}B.
However, Fig \ref{fig:daily_volume}B also shows that the trading volume of
companies which announce their earning reports AfterClose (blue line)
is considerably higher than for the BeforeOpen announcements (red line).
This is not due to their higher capitalization, which is only slightly
above the average (see Table \ref{tab:tweets}).
One possible explanation for the increased trades can be the assimilation 
hypothesis \cite{doyle2009timing}: AfterClose reporting allows the market 
more time to assimilate the information in the announcement.
Companies that announce their earnings AfterClose are typically
more technologically oriented, e.g., Microsoft, IBM, Cisco, Intel, and
have more complex operations.

\begin{figure}[!h]
\centering
\includegraphics[width=\textwidth]{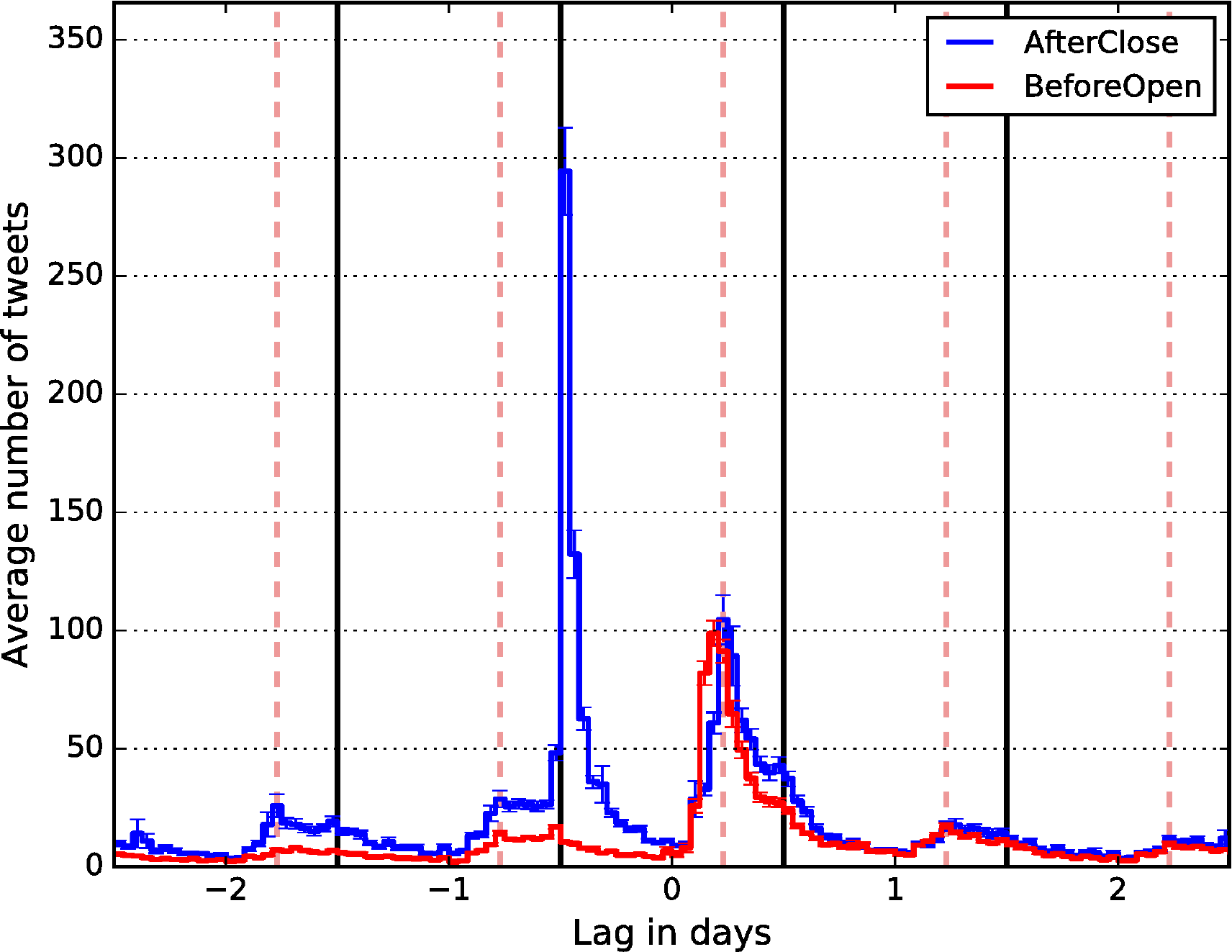}
\caption{{\bf Hourly distribution of tweets around the Earnings Announcements.}
Day 0 is the day of the EAs.
Dashed lines denote market open (9:30 a.m. US/Eastern) and solid lines 
denote market close (4:00 p.m. US/Eastern). Solid lines also
delimit days for aggregation of tweets at the daily resolution.
Error bars denote one standard error.}
\label{fig:hourly_dist}
\end{figure}

Next we compare the Twitter volumes between the companies
which make the AfterClose and BeforeOpen announcements.
Fig \ref{fig:hourly_dist} shows the average hourly number of tweets
around the EAs (day $0$) for both types of companies.

The comparison of the Twitter volumes at day $0$ shows a considerably higher
average number of tweets for the AfterClose announcements, consistent with
the higher trading activity.
Hourly distribution of tweets also matches the timing of the announcements:
the AfterClose peak is immediately after the market on day $-1$ closes and
the announcement is made. BeforeOpen peaks immediately before the market
opens on day 0. In both cases, the Twitter activity is very similar after
the market opens on day 0.

\subsection{Event study applications}

The Earnings Announcements are important events which trigger higher trading
on stock exchange and also draw attention and comments on social media.
Is there also any correspondence between the stance of Twitter users and
abnormal returns of the stocks after the EA events? The goal of this
subsection is to answer this question by applying the event study methodology.

An event study captures the impact of external events on the stock returns.
In an event study, Cumulative Abnormal Returns (CAR) are defined as a measure 
of returns which are above or below the overall market returns.
Details of the event study methodology applied here are in the
``\nameref{subsect:event_study}'' subsection in ``\nameref{sec:methods}''.

External events that we consider in this paper are EAs only. Their dates are known
in advance, and we compute their polarity from the Twitter sentiment.
Details about how the sentiment of the tweets on the day of the EA (or the day before)
is used to derive the polarity of the event (negative, neutral, or positive) are
in the ``\nameref{subsect:cat_EA}'' subsection in ``\nameref{sec:methods}''.

The event window in this study starts on the day before the EA (day $-1$)
and runs until 10 days after the EA. For each day, we check
the correspondence between the polarity of the events and the direction of CARs
(profit or loss), significance of CARs, and their magnitude.
The null hypothesis $H_{0}$ is that the EAs have no impact on the CARs.
We test the $H_{0}$ for the AfterClose and BeforeOpen announcements
separately.

\begin{figure}[!h]
\centering
\includegraphics[width=\textwidth]{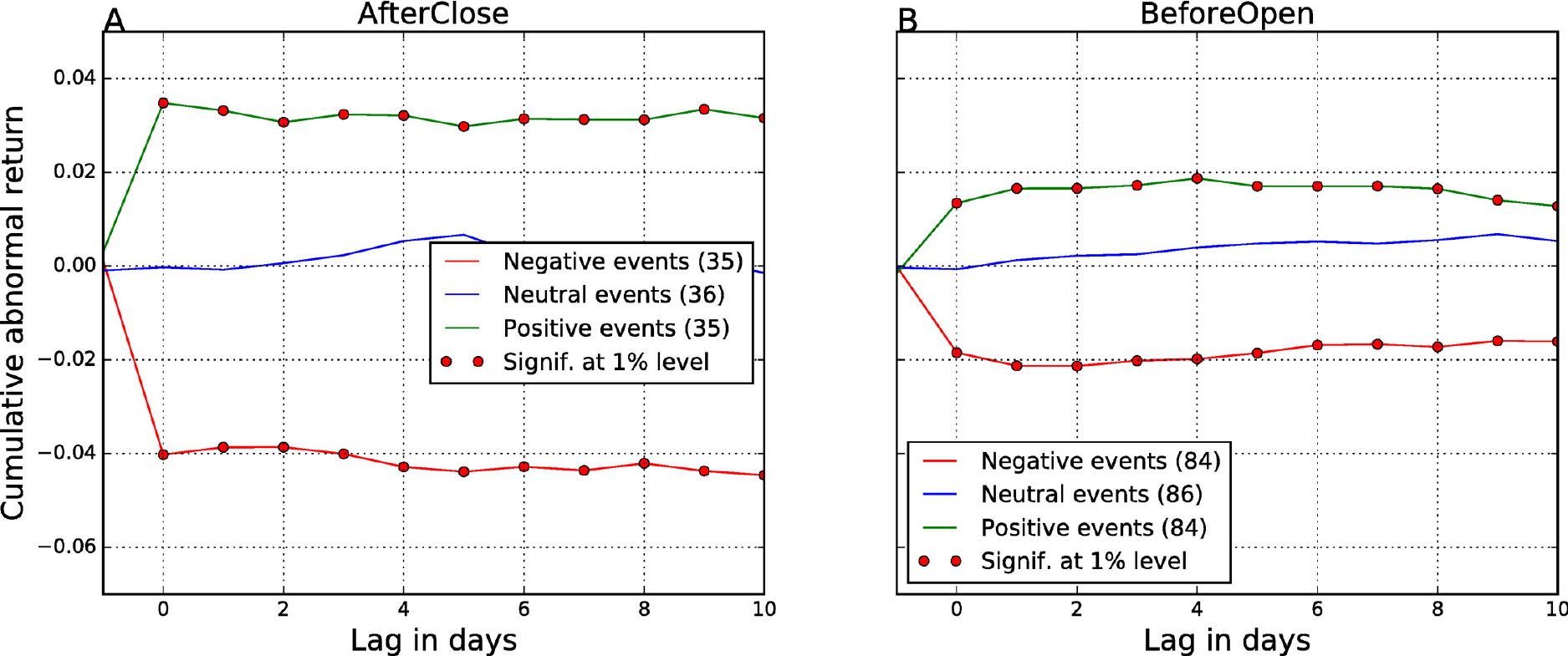}
\caption{{\bf Cumulative Abnormal Returns---polarity of the EAs is computed from 
the sentiment of tweets on day~0.} 
The AfterClose (A) and BeforeOpen (B) events are analyzed separately.
Different line colors denote different polarity of the events:
green line denotes positive events, blue line neutral events, and red
line negative events. In the legends, the numbers in parentheses are the
numbers of different types of events.
Days when CARs are significant at the 1\% level are marked with red dots.}
\label{fig:car_s0}
\end{figure}

Fig \ref{fig:car_s0} gives the results for the EA events, when we determine
their polarity from the Twitter sentiment on the day of the EA, \sentzero.
Results show that the null hypothesis $H_{0}$ is rejected for all the days
after the EA.
There is almost a perfect match between the polarity of the EAs, determined
from the Twitter sentiment, and the direction of CARs.
Neutral announcements (blue lines) yield no returns (CARs are around zero), 
while positive (green lines) and negative announcements (red lines) are aligned 
with profits (positive CARs) and losses (negative CARs), respectively. 
The magnitude of CARs is high (around 2--4\%),
and all of them are significant at the 1\% level (denoted by red dots).
These results are consistent with the existing literature on the information
contents of the EA reports. In our previous event study
\cite{Ranco2015eventstudy}, where we analyzed all Twitter peaks as events and
not just EAs, the CARs were between 1--2\%. Here, where we have longer time period
and the EA events only, the CARs are between 2--4\%.
This confirms that the Twitter sentiment correctly captures the contents of the
EA reports.

However, the magnitude of CARs is different for the AfterClose 
(around 4\% in Fig \ref{fig:car_s0}A) and BeforeOpen events (around 2\% in
Fig \ref{fig:car_s0}B). Also, the CARs for BeforeOpen are declining with time,
and the neutral line shows a slight upward trend. This suggests that tweets convey
a weaker signal for the BeforeOpen announcements, in addition or due to their
lower volume, as compared to the AfterClose announcements.

\begin{figure}[!h]
\centering
\includegraphics[width=\textwidth]{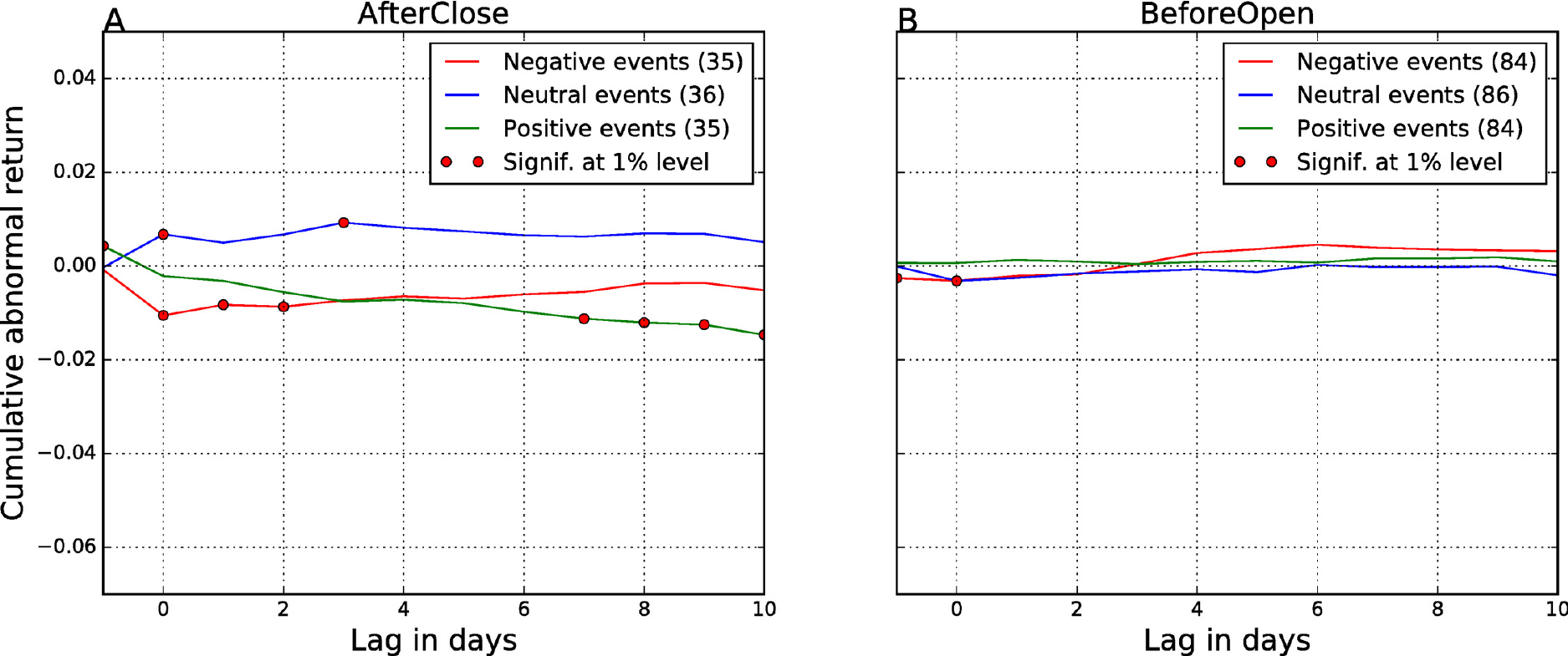}
\caption{{\bf Cumulative Abnormal Returns---polarity of the EAs is computed from 
the sentiment of tweets on day~$-1$.} 
The AfterClose (A) and BeforeOpen (B) events are analyzed separately.
Different line colors denote different polarity of the events:
green line denotes positive events, blue line neutral events, and red
line negative events. In the legends, the numbers in parentheses are the
numbers of different types of events.
Days when CARs are significant at the 1\% level are marked with red dots.}
\label{fig:car_s_1}
\end{figure}

Next, we investigate if there is any anticipation of information about the
upcoming EAs in the social media. We determine the polarity of the EA events
from tweets on the day before the announcement (day $-1$). All other parameters
of the events study remain the same. The results are in Fig \ref{fig:car_s_1}.

In Fig \ref{fig:car_s_1}B we see that the returns are practically zero, for all
types of EAs. We can conclude that there is no information about the BeforeOpen 
announcements in the Twitter posts on the day before the EA.

The returns in Fig \ref{fig:car_s_1}A are small, but nonzero. 
However, the polarity of the neutral and positive events
from the tweets does not match the sign of the corresponding CARs
(green and blue lines in Fig \ref{fig:car_s_1}A are misplaced).
There is a weak signal for the negative AfterClose events
(red line in Fig \ref{fig:car_s_1}A).
The negative CARs for the first three days after the EAs are small
(about 1\%), but statistically significant (marked by red dots).
We exploit this result in the next subsection where we design a trading strategy.

It is important to note the impact of different alignments between the
Twitter data and the EAs on the predictive power of the tweets.
If the tweets are delimited at calendar days, one might observe a
spurious predictive power of the Twitter sentiment.
For the AfterClose announcements, there is a peak of Twitter activity
immediately after the market closes, but before midnight 
(see Fig \ref{fig:hourly_dist}). If this is aligned with trading on
day $-1$, and not on day 0, one might well observe the results similar
to Fig \ref{fig:car_s0}A. Then one can draw a misleading conclusion 
that the Twitter sentiment on day $-1$ anticipates significant CARs
on day 0 and subsequent days. This problem was already identified in the 
financial literature \cite{berkman2009event}, and here we reiterate its
proper treatment in the social media context.

\subsection{Exploring trade returns}

The goal of this subsection is to develop an actionable trading strategy
based on the Twitter sentiment.
The results of the event study, in terms of the CARs,
cannot be directly exploited for trading. 
They show that our results, obtained with automated Twitter sentiment 
classification, are consistent with the existing financial literature.
For trading, however, they provide just some hints
on the timings and polarity of the EAs worth exploring.
A trading strategy has to specify which stocks to select and when to
buy/sell them. The aim of this subsection is to modify the event study
to identify the stocks (from the type and polarity of the EAs), the
actions (buy or sell, from the computed returns), and exact timings
of the trades (from the time line of the returns).

Here we analyze trade returns, as defined in 
Eq (\ref{eq:trade_return_trading}), instead of CARs.
We assume that one buys/sells a stock at a closing price of the day before 
the EA (day $-1$), and then sells/buys the same stock $d$ days after the EA.
For comparison, we also show the results when one buys/sells the DJIA index,
instead of the individual stock. This should yield results very similar to the
event study, which already incorporates a market model in the CARs.

\begin{figure}[!h]
\centering
\includegraphics[width=\textwidth]{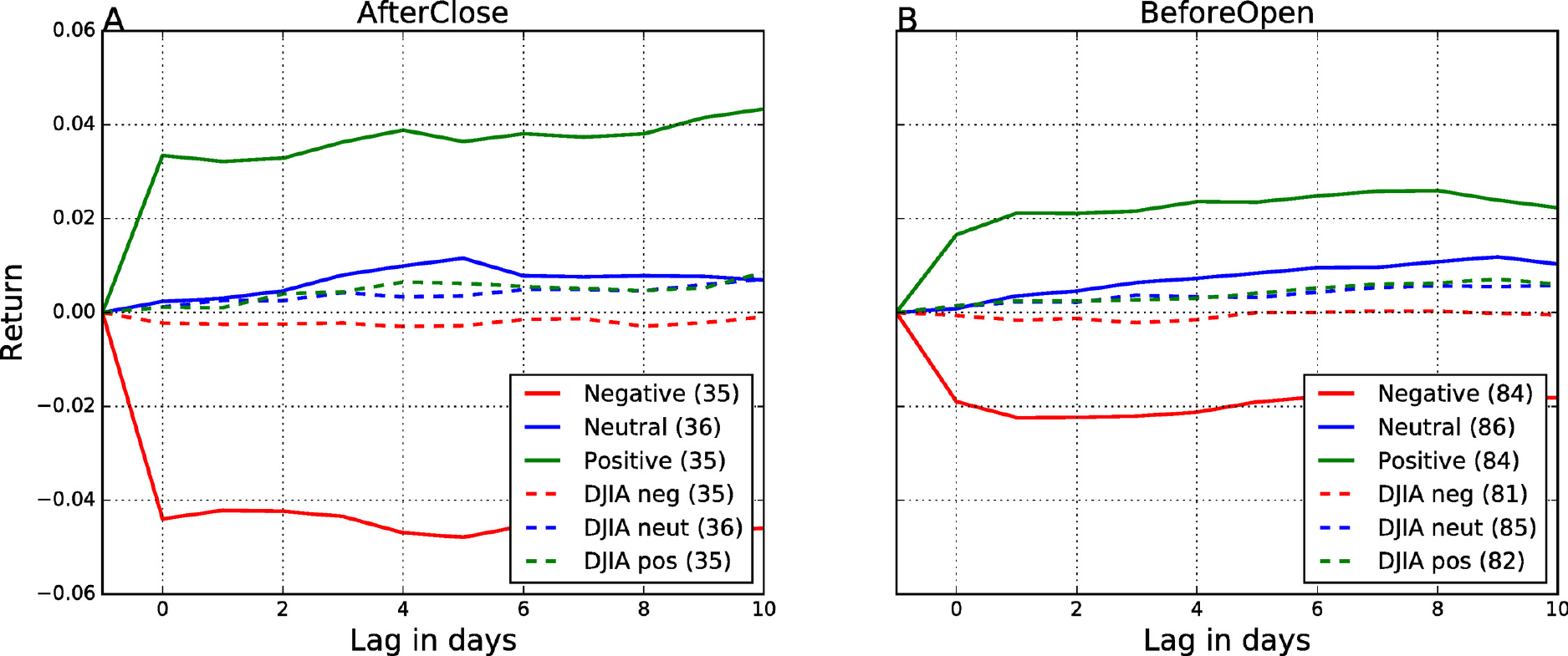}
\caption{{\bf Trade returns---polarity of the EAs is computed from tweets on day~0.} 
The AfterClose (A) and BeforeOpen (B) events are analyzed separately.
Solid lines denote trades with individual stocks, and dashed lines
denote the corresponding trades with the DJIA index.
Line colors denote different polarity of events as determined from
the sentiment of tweets.}
\label{fig:return_nowcast}
\end{figure}

Our starting point are the results of the event study.
We first make an unrealistic
assumption that at the trade on day $-1$ we already foresee the Twitter sentiment 
on the next day, \sentzero. We therefore sell/hold/buy a stock on day $-1$ 
if \sentzero\, is negative/neutral/positive, respectively. 
The results are in Fig \ref{fig:return_nowcast}.
The returns are high, as expected, and very similar to CARs in
Fig \ref{fig:car_s0}.

\begin{figure}[!h]
\centering
\includegraphics[width=\textwidth]{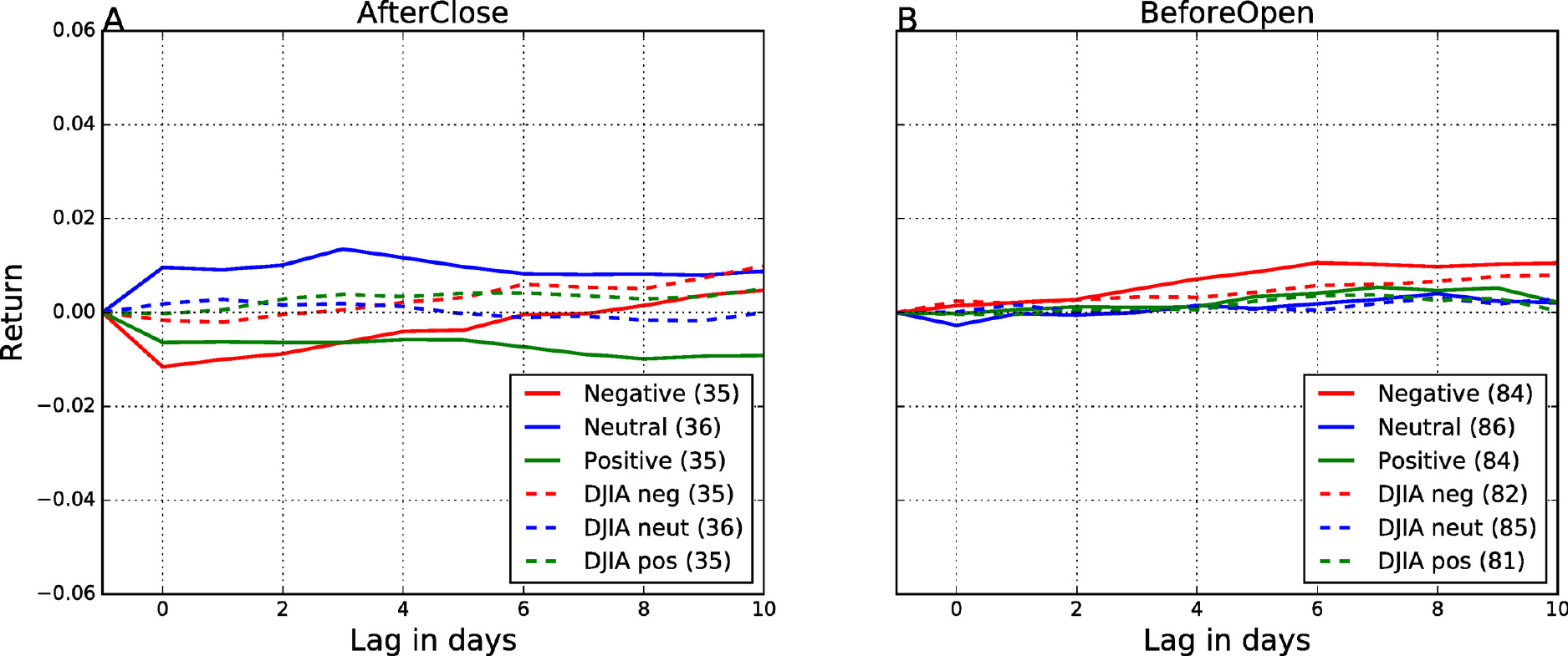}
\caption{{\bf Trade returns---polarity of the EAs is computed from tweets on day~$-1$.} 
The AfterClose (A) and BeforeOpen (B) events are analyzed separately.
Solid lines denote trades with individual stocks, and dashed lines
denote the corresponding trades with the DJIA index.
Line colors denote different polarity of events as determined from
the sentiment of tweets.}
\label{fig:return_forecast}
\end{figure}

Next, we make a realistic assumptions, and trade on day $-1$, based on the
tweets and sentiment of the same day, \sentone.
The results are in Fig \ref{fig:return_forecast}, again very similar to
CARs in Fig \ref{fig:car_s_1}.
They show that the polarity of the BeforeOpen announcements cannot be
predicted one day in advance from the tweets alone (Fig \ref{fig:return_forecast}B).
For the AfterClose announcements, some low return can be expected only for
the negative events, and is already diminishing after the announcement day
(red line in Fig \ref{fig:return_forecast}A).
A possible explanation is that companies leak positive news already 
several days in advance, but they hold negative news for as long as 
they can \cite{Sprenger2014jbfa}.

The above results provide the guidelines on how to devise a trading strategy.
We can only trade based on the Twitter sentiment before the EAs, therefore
the returns in Fig \ref{fig:return_forecast} are relevant.
The magnitude of returns is around 1\% only for the AfterClose announcements
(Fig \ref{fig:return_forecast}A) and the polarity of the Twitter sentiment
is aligned only with the negative returns (red line in Fig \ref{fig:return_forecast}A).
Based on these insights we can devise a simple trading strategy:
\begin{itemize}
\item consider only the AfterClose announcements,
\item trade only on negative events, i.e., polarity of \sentone\, is negative,
\item sell (short) a stock at day $-1$, and buy it back at day 0.
\end{itemize}
Note that short selling is a common practice of selling a stock that
is not currently owned.

We evaluate this simple trading strategy by backtesting it on three and half years 
worth of historical data, from June 1, 2013 until December 31, 2016.
We assume that all the trades are executed at the closing 
price, and that all returns are reinvested.
We also take into account spread, chosen conservatively at 0.05\$ per share. 
In practice, spread is usually around one cent or less \cite{nysedata_factbook}.

\begin{figure}[!h]
\centering
\includegraphics[width=\textwidth]{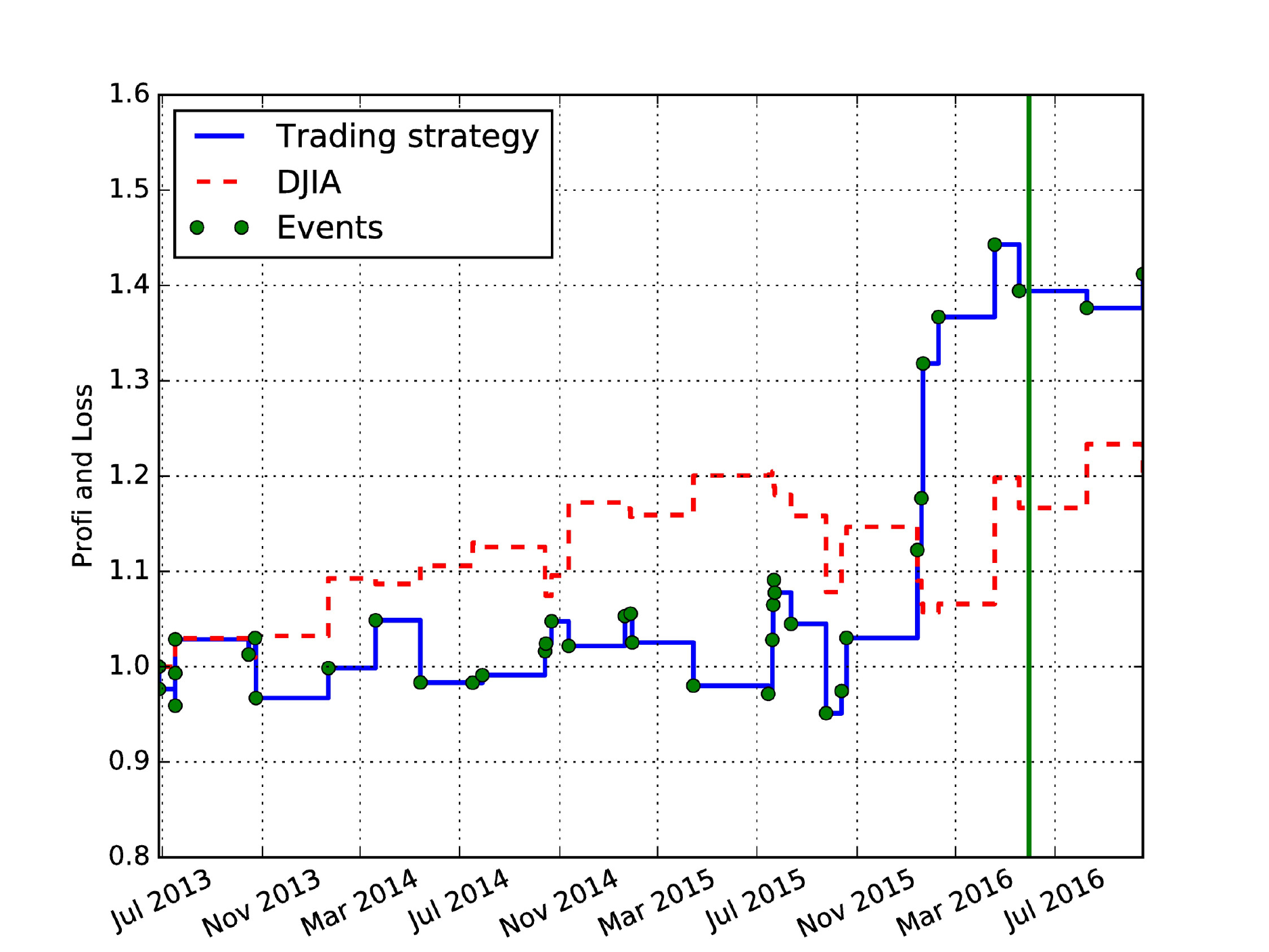}
\caption{{\bf Trading simulation based on the simple trading strategy.}
Blue line shows the trade returns, and green dots denote the 37 negative EA events that
triggered the trades. For comparison, the red dashed line shows the
value of the DJIA index. The green vertical line delimits the first three
years of data from the last half year of new data.}
\label{fig:trading_strategy}
\end{figure}

The result of the trading simulation are in Fig \ref{fig:trading_strategy}. 
The green dots represent the negative EA events (i.e., \sentone\, was negative) 
during which our strategy executed a trade. 
Solid blue line represents a cumulative return of our strategy, 
assuming an initial investment of $1.0$. 
Dashed red line shows the return of the DJIA index, considered as benchmark.
The green vertical line delimits the first three years of data, on which
the event study was applied, from the last half year of new data.

The simple trading strategy executes 37 trades (short sells and repurchases),
and yields a 42\% return in the tree and half years. For comparison,
the DJIA index gained about 21\% in the same period.
While this simple strategy considerably outperforms the benchmark, 
most of the difference in the returns was realized in a half a year,
from January to June 2016.
The six EAs in this period, classified as negative and yielding most of the profits, 
are from the following companies: IBM (twice), INTC, AXP, DIS and CSCO.
The above trading strategy was derived from the three years of data.
We have tested the same strategy also on the new data, from July to December 2016.
In this period, there are only two trades (with DIS and IBM) and the
profits are negligible.

From these results we cannot draw any reliable conclusion about the performance
of the proposed trading strategy. We do not claim that the relatively
large returns are significant, nor that the strategy would yield similar
results in the future.
However, it is interesting to note that this trading strategy exhibits 
low profits and losses during the first two and a half years, and that
it does not follow the index.

\subsection{Relation between the Twitter sentiment and earnings surprise}

The goal of this section is to compare the information about the
Earnings Announcements extracted from social media to financial expectations.
Financial analysts estimate earnings per share of a company a few weeks in advance 
of its EA (see e.g., \url{http://www.zacks.com/earnings/}).
This estimate is more or less accurate, but raises some expectations.
When the actual, reported earnings are different, the result is an
earnings surprise, negative or positive.
Earnings surprise (\textit{ES}) is defined as a normalized difference between 
the reported and estimated earnings of a company, 
see Eq (\ref{eq:eps}).
Earnings surprise is often used in event studies to categorize the EA events. 
For example, in the original event study \cite{mackinlay1997event},
MacKinlay proposes to categorize the EA event as positive if the actual earnings
exceed the expected by more than 2.5\% (and the opposite for the negative events).
In our study we use the Twitter sentiment to categorize the EA events.

\begin{figure}[!h]
\centering
\includegraphics[width=\textwidth]{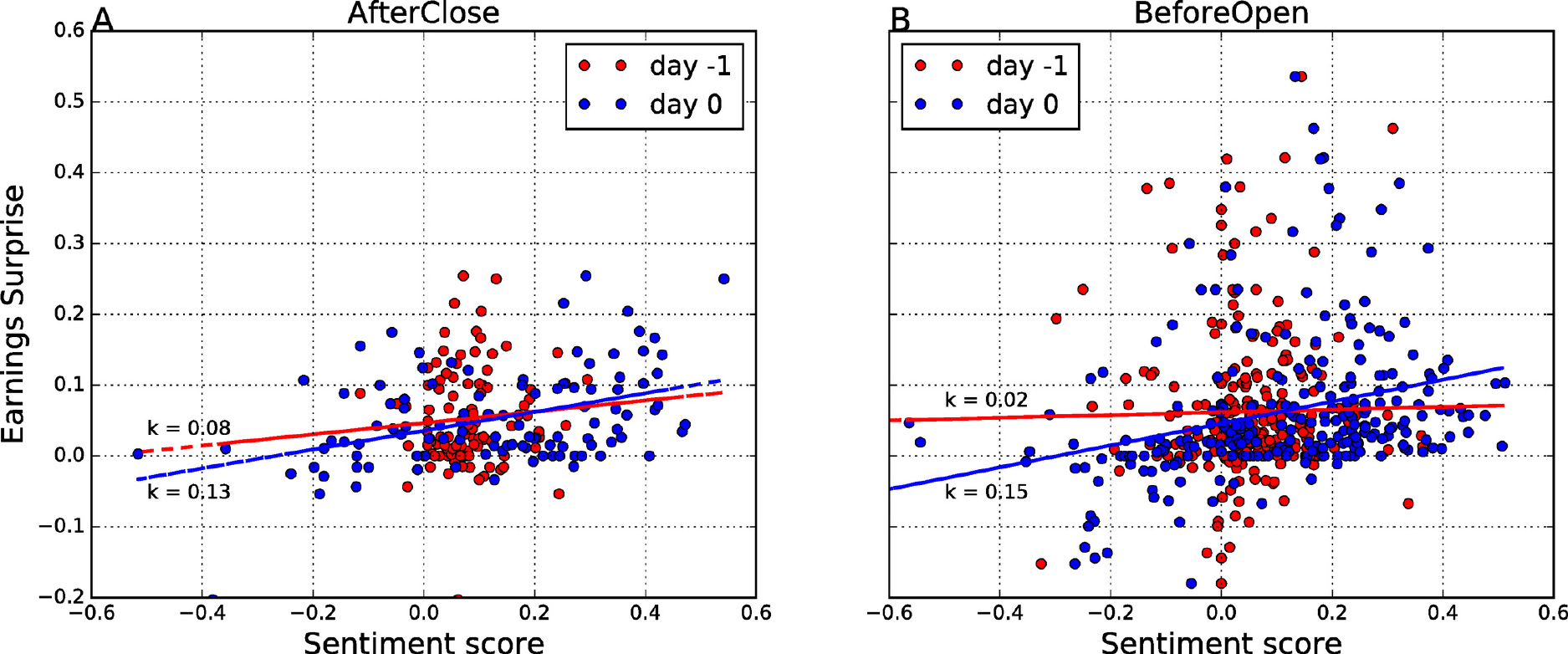}
\caption{{\bf Relation between the sentiment score and earnings surprise.}
The AfterClose (A) and BeforeOpen (B) events are analyzed separately.
Blue dots denote the polarity of the events on day $-1$, and red dots on day 0.
The corresponding regressions are represented by solid lines.}
\label{fig:surACBO}
\end{figure}

We compare the information contents of the Twitter sentiment to the
earnings surprise. We apply the ordinary least squares estimate to
determine the linear regression between the sentiment score and \textit{ES}.
The results are shown in Fig \ref{fig:surACBO}. 
The regression lines have the following form:

\begin{tabular}{rl}
AfterClose (day 0):    & $ES = 0.13 \cdot Sent(0) + 0.04 \;\; (R^2 = 0.17)$   \;, \\
BeforeOpen (day 0):    & $ES = 0.15 \cdot Sent(0) + 0.05 \;\; (R^2 = 0.09)$   \;, \\
AfterClose (day $-1$): & $ES = 0.08 \cdot Sent(-1) + 0.05 \;\; (R^2 = 0.007)$ \;, \\
BeforeOpen (day $-1$): & $ES = 0.02 \cdot Sent(-1) + 0.06 \;\; (R^2 = 0.0003)$ \;. \\[2ex]
\end{tabular}

The results of the linear regression suggest that the sentiment score of day 0 and 
earnings surprise are related, but very weakly.
However, no evidence of relation is found between $Sent(-1)$ and $ES$, since the $R^2$ 
coefficient shows that $Sent(-1)$ explains less than 1\% of the total variance.
In more detail, the linear models which use $Sent(-1)$ have very small 
explanatory power, for AfterClose as well as BeforeOpen EAs.

\section{Conclusions}

The present study shows that there is a considerable interplay between
the social media and stock market. To some extent the results of the related work
are corroborated, but we also present more detailed, in-depth analyses.
In particular, we focus only on the Earnings Announcements, the events that
draw the highest trading activity and social media attention.
We find important differences regarding the timings of the announcements:
before the market opens versus after the market closes. These differences
have to be taken into account when aggregating the Twitter data at
the daily resolution and when aligning the Twitter and market data.

We applied the event study methodology, where the Twitter sentiment determines
the polarity of the Earnings Announcement reports. We show that the Twitter sentiment
is a very good interpreter of the announcements contents. Cumulative Abnormal Returns
are high and statistically significant. However, we did not find evidence that
the Twitter sentiment alone can predict the returns one day before they are announced.
This negative result might be due to the chosen alignment between the Twitter
and market data for the after-hours announcements. If the data are not aligned as
recommended in the literature, one might observe spurious predictive impact of
the Twitter sentiment on price returns.

We also analyze earnings surprise, which is a measure frequently used in event studies. 
Our comparison to the Twitter sentiment shows that they have little in common. 
A possible reason might be that the aggregate measure from social media contains 
different information than the aggregated anticipations of the financial analysts.

This study is limited to Earnings Announcements only, where we observe considerably 
elevated trading and tweeting activities.
There are other, unexpected events which can be identified with peaks in
social media activities. These events can have significant impact on the market, 
and the Twitter sentiment can play an important role
in devising social media-enhanced trading strategies.


\section{Methods}
\label{sec:methods}

In this section we first outline our Twitter sentiment classification approach.
We then show how to determine the polarity of the EA events from
the sentiment of tweets on a particular day.
The polarity of the events is then used in the event study.
The event study methodology is briefly summarized in the last subsection.

\subsection{Sentiment classification}
\label{subsect:sent_class}

All the collected financial tweets are labeled with sentiment.
The sentiment captures the leaning or stance of a Twitter user
with respect to the anticipated future move of the stock.
A stock mentioned in the tweet is identified by a cash-tag (e.g., ``\$IBM'').
The anticipated change of its price is
approximated by three sentiment values: negative (stock price will go down), 
neutral (price will remain unchanged), or positive (stock price will go up).
The labeling of tweets is automatic, by applying a
sentiment classification model.

Our approach to automatic sentiment classification is based
on supervised machine learning. The procedure consists of the following steps:
(i) a large sample of tweets (about 100,000) is first manually annotated with stance
by financial experts,
(ii) the labeled set is used to train and tune a classifier,
(iii) the classifier is evaluated by cross-validation and
compared to the inter-annotator agreement, and
(iv) the classifier is applied to the whole set of collected tweets.

There are many supervised machine learning algorithms suitable for
training a sentiment classifier. 
Often, variants of Support Vector Machine (SVM) \cite{Vapnik1995}
are used, because they are well suited for large scale text categorization
tasks, are robust, and perform well.
For this study, a two plane SVM classifier was constructed \cite{Mozetic2016annot}.
The two plane SVM assumes the ordering of sentiment values and
implements ordinal classification. 
It consists of two SVM classifiers: 
One classifier is trained to separate the negative tweets from the
neutral-or-positives; the other separates the negative-or-neutrals 
from the positives.
The result is a classifier with two hyperplanes that partitions the 
vector space into three
subspaces: negative, neutral, and positive. During classification,
the distances from both hyperplanes determine the predicted sentiment value.


The labeled tweets for each stock are aggregated on a daily basis, and
the sentiment score, defined in Eq (\ref{eq:sentiment_score}), is computed.
Note that in our previous event study research \cite{Ranco2015eventstudy}, 
we operated with sentiment polarity instead of the sentiment score.
Sentiment polarity ignores the neutral tweets, and is defined as
$
\frac{N_d(pos) - N_d(neg)}{N_d(pos) + N_d(neg)} .
$

\subsection{Polarity of the EA events}
\label{subsect:cat_EA}

\begin{figure}[!h]
\centering
\includegraphics[width=\textwidth]{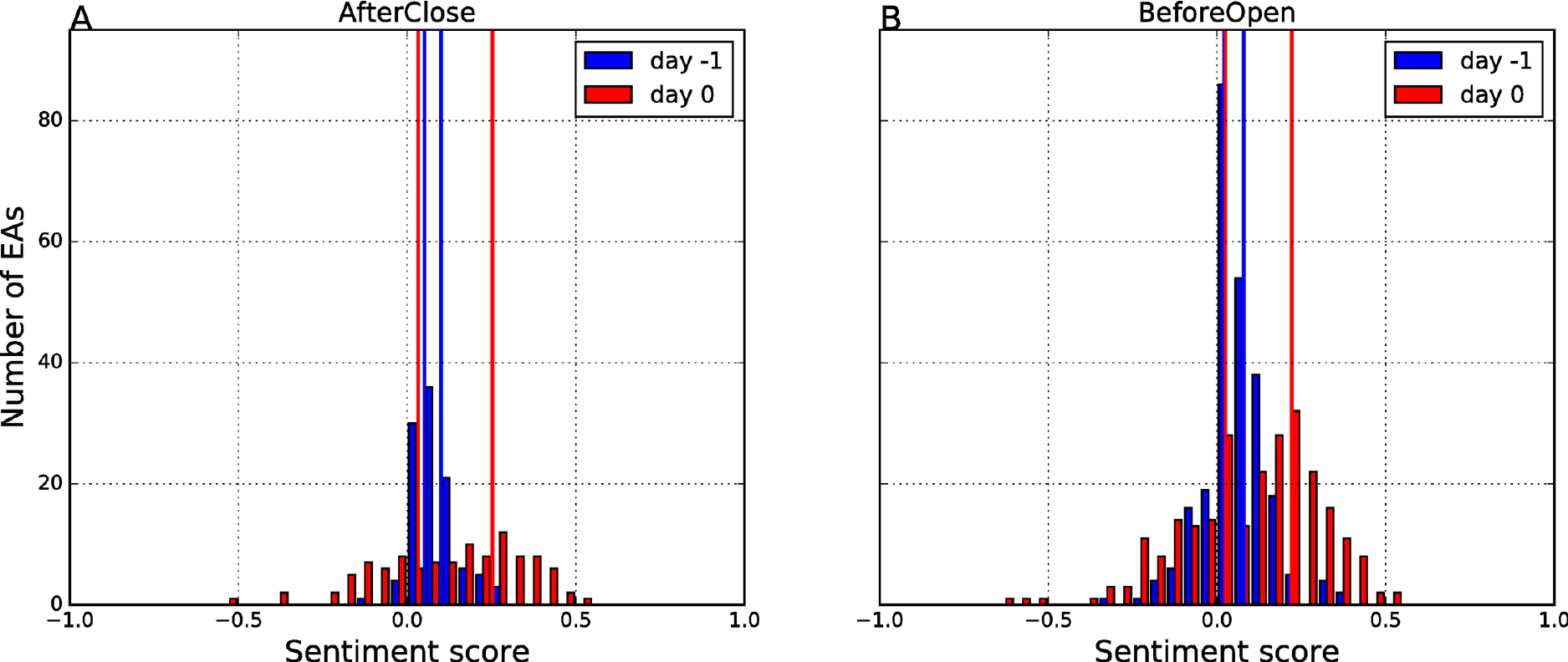}
\caption{{\bf Sentiment distribution of all the Earnings Announcements.}
Sentiment score is computed from the tweets on day $-1$ (blue) and day 0 (red),
separately for the AfterClose (A) and BeforeOpen (B) events.
The vertical lines mark the thresholds used to discriminate between the 
negative, neutral, and positive event polarity.}
\label{fig:sent_dist}
\end{figure}

The event study methodology requires that external events, EAs in our case,
are categorized whether they should have negative, positive, or no effect 
on stock returns.
We determine the polarity of the EA events from the sentiment scores,
aggregated on a particular day (day 0 or $-1$).
Distributions of the sentiment scores, for all the EAs, for the two relevant
days, are in Fig \ref{fig:sent_dist}.

Note that the number of the AfterClose events (Fig \ref{fig:sent_dist}A) is 106,
and the number of the BeforeOpen events (Fig \ref{fig:sent_dist}B) is 253.
The sentiment scores, \sentzero\, (red bars) and \sentone\, (blue bars)
are roughly normally distributed with slightly positive means.
Note also that the strength of the Twitter sentiment signal on day $-1$
is much weaker than on day 0. This is indicated by the large number of events
with values of \sentone\, close to 0. In other words, the tweets of day $-1$ contain 
less information about the upcoming event than the tweets of day 0, as expected.

We determine the polarity of the EAs from the distribution of \sentzero\,
and \sentone. We define thresholds in such a way, that the three categories are 
distributed uniformly.
The thresholds are reported in Table \ref{tab:thresholds}, and are also shown 
in Fig \ref{fig:sent_dist}, as red and blue vertical lines. 

\begin{table}[!h]
\centering
\caption{Categorization of the EA events from the sentiment scores on days 0 and $-1$.}
\begin{tabular}{l|ll|ll}
\hline
EA event & \multicolumn{2}{c|}{AfterClose} & \multicolumn{2}{c}{BeforeOpen} \\
polarity & day $0$           & day $-1$         & day $0$           & day $-1$ \\
\hline
Negative & $Sent(0) \in (-1, 0.03]$   & $Sent(-1) \in (-1, 0.05]$   & $Sent(0) \in (-1, 0.02]$   & $Sent(-1) \in (-1, 0.02]$ \\
Neutral  & $Sent(0) \in (0.03, 0.25]$ & $Sent(-1) \in (0.05, 0.10]$ & $Sent(0) \in (0.02, 0.22]$ & $Sent(-1) \in (0.02, 0.08]$ \\
Positive & $Sent(0) \in (0.25, 1)$    & $Sent(-1) \in (0.10, 1)$    & $Sent(0) \in (0.22, 1)$    & $Sent(-1) \in (0.08, 1)$ \\
\hline
\end{tabular}
\label{tab:thresholds}
\end{table}


Putting a threshold on a signal is always somehow arbitrary, and there is no 
systematic treatment of this issue in the event study \cite{mackinlay1997event}. 
The justification for our approach, already used in \cite{Ranco2015eventstudy},
is that sentiment should be regarded in relative terms, in the context of related events.
Sentiment score has no absolute meaning, but provides just an 
ordering of events on the scale from $-1$ (negative) to $+1$ (positive).
The most straightforward choice is to distribute all the events
uniformly between the three classes.
In the closely related work by Sprenger et al. \cite{Sprenger2014jbfa}, 
the authors use the percentage of positive tweets for 
a given day $d$, to determine the event polarity. 
Since they also report an excess of positive tweets,
they use the median share of positive tweets as a threshold between 
the positive and negative events.

\subsection{Event study methodology}
\label{subsect:event_study}

The event study methodology was originaly defined
in financial econometrics \cite{mackinlay1997event, campbell1997econometrics}.
The first adaptations and applications to social media data were reported
by Sprenger et al. \cite{Sprenger2014jbfa} and Ranco et al. \cite{Ranco2015eventstudy}.
In the current study, there are two differences with respect to our
previous work \cite{Ranco2015eventstudy}: here we focus just on the events anticipated
in advance (i.e., EAs), and we use a shorter event window, compatible with
the devised trading strategy.

In summary, in the current event study we use an event window of 12 trading days, 
i.e., one day before the EA event, and up to 10 days afterwards. 
We use an estimation window of 120 trading days, and a market model as 
the normal performance model, estimated with an 
ordinary least squares regression of the DJIA returns. 

\subsubsection{Cumulative abnormal returns}

An event study is based on the premise that in order to correctly capture the impact 
of an event, a measure of abnormal price return must be defined. 
This measure is the actual price return minus the normal return of the stock 
during the event window. For company $i$ and event date $d$ the abnormal return is:
\begin{equation}
AR_{i,d}=R_{i,d}- E[R_{i,d}]
\end{equation}
where $AR_{i,d}$, $R_{i,d}$, $E[R_{i,d}]$ are the abnormal, actual, and expected 
normal returns, respectively. The normal performance model used in this work is 
the market model: it assumes a linear relation between the overall market return 
and the return of the stock. More details are given in \cite{Ranco2015eventstudy}.

In order to draw overall conclusions for the set of events being analyzed, 
the abnormal return observations must be first aggregated. 
The aggregation is performed through time and across stocks.
By aggregating across all the stocks, we obtain:
\begin{equation}						
\overline{AR}_\tau=(1/N)\sum_{i=1}^{N}AR_{i,\tau} \; .
\end{equation}
The cumulative abnormal return ($CAR$) from time $\tau_1$ to $\tau_2$ is 
the sum of the abnormal returns: 
\begin{equation}
CAR(\tau_1,\tau_2) = \sum_{\tau=\tau_1}^{\tau_2}\overline{AR}_\tau \; .
\end{equation}

For the calculation of the $CAR$ variance, we assume that 
$\sigma_{AR}^2=\sigma_{\epsilon_ {i,t}}^2$  (shown in \cite{mackinlay1997event}):
\begin{equation}
var(CAR(\tau_1,\tau_2))= (1/N^2)\sum_{i=1}^{N}(\tau_2 - \tau_1 +1)\sigma_{\epsilon_i}^2 
\end{equation}
where $N$ is the total number of events.
Finally, we introduce the test statistic $\hat{\theta}$. 
This quantity is used to assess whether the impact of the external event 
on the cumulative abnormal returns is significant. The test statistic is defined as:
\begin{equation}
\frac{CAR(\tau_1,\tau_2)}{\sqrt[2]{var(CAR(\tau_1,\tau_2))}} =
\hat{\theta} \thicksim \mathcal{N}(0,1) 
\end{equation}
where $\tau$ is the time index inside the event window, 
and $|\tau_2 - \tau_1|$ is the total length of the event window.

\section*{Acknowledgments}
The authors would like to thank Sowa Labs (\url{http://www.sowalabs.com})
for providing the sentiment labeled Twitter data about the 30 DJIA
companies, and Sebastian Schroff for valuable insights
regarding the timing of Earnings Announcements.

This work was supported in part by the EC projects SIMPOL (no. 610704) and DOLFINS 
(no. 640772), and by the Slovenian ARRS programme Knowledge Technologies (no. P2-103).


\end{document}